\documentclass{aa}  

\usepackage{graphicx}
\usepackage{txfonts}
\usepackage[]{hyperref} 
\usepackage{booktabs} 
\usepackage{array} 
\usepackage{paralist} 
\usepackage{verbatim} 
\usepackage{ulem}
\usepackage{xspace}
\usepackage{siunitx}
\usepackage{amssymb}
\usepackage{capt-of}
\usepackage{natbib}
\bibpunct{(}{)}{;}{a}{}{,} 

\graphicspath{{./img/}}

\makeatletter
\renewcommand*\aa@pageof{, page \thepage{} of \pageref*{LastPage}}
\makeatother

\addto\extrasenglish{%
}

\begin{document}

   \title{Retrieval of the dayside atmosphere of WASP-43b with CRIRES$^+$}

   \subtitle{}

   \author{
            F. Lesjak \inst{1} \and 
            L. Nortmann \inst{1} \and
            F. Yan \inst{2} \and
            D. Cont \inst{3,4,1}  \and
            A. Reiners\inst{1} \and
            N. Piskunov\inst{5} \and
            A. Hatzes\inst{6} \and
            L. Boldt-Christmas\inst{5}\and
            S. Czesla\inst{6} \and
            U. Heiter\inst{5} \and
            O. Kochukhov\inst{5} \and
            A. Lavail\inst{7,5} \and
            E. Nagel\inst{1,8,6} \and
            A. D. Rains\inst{5} \and
            M. Rengel\inst{9} \and
            F. Rodler\inst{10} \and
            U. Seemann\inst{11, 1}\and
            D. Shulyak \inst{12}
          }

   \institute{
            Institut f\"ur Astrophysik und Geophysik, Georg-August-Universit\"at, Friedrich-Hund-Platz 1, 37077 G\"ottingen, Germany \\
            email: fabio.lesjak@uni-goettingen.de\and
            Department of Astronomy, University of Science and Technology of China, Hefei 230026, China \and
            Universitäts-Sternwarte, Fakultät für Physik, Ludwig-Maximilians-Universität München, Scheinerstr. 1, 81679 München, Germany \and
            Exzellenzcluster Origins, Boltzmannstraße 2, 85748 Garching, Germany \and
            Department of Physics and Astronomy, Uppsala University, Box 516, 75120 Uppsala, Sweden \and
            Thüringer Landessternwarte Tautenburg, Sternwarte 5, 07778 Tautenburg, Germany \and
            Institut de Recherche en Astrophysique et Planétologie, Université de Toulouse, CNRS, IRAP/UMR 5277, 14 avenue Edouard Belin, F-31400, Toulouse, France \and
            Hamburger Sternwarte, Universität Hamburg, Gojenbergsweg 112, 21029 Hamburg, Germany \and
            Max-Planck-Institut für Sonnensystemforschung, Justus-von-Liebig-Weg 3, 37077 Göttingen, Germany \and
            European Southern Observatory, Alonso de Cordova 3107, Vitacura, Santiago, Chile \and
            European Southern Observatory, Karl-Schwarzschild-Str. 2, 85748 Garching bei München, Germany \and
Instituto de Astrofísica de Andalucía - CSIC, Glorieta de la Astronomía s/n, 18008 Granada, Spain
             }

   \date{Received June 11, 2023; accepted July 17, 2023}

    \abstract{Accurately estimating the C/O ratio of hot Jupiter atmospheres is a promising pathway towards understanding planet formation and migration, as well as the formation of clouds and the overall atmospheric composition. The atmosphere of the hot Jupiter WASP-43b  
has been extensively analysed using low-resolution observations with HST and Spitzer, but these previous observations did not cover the K band, which hosts prominent spectral features of major carbon-bearing species such as CO and CH$_{4}$. As a result, the ability to establish precise constraints on the C/O ratio was limited. Moreover, the planet has not been studied at high spectral resolution, which can provide insights into the atmospheric dynamics. 

In this study, we present the first high-resolution dayside spectra of WASP-43b with the new CRIRES$^+$ spectrograph. By observing the planet in the K band, we successfully detected the presence of CO and provide evidence for the existence of H$_2$O using the cross-correlation method. This discovery represents the first direct detection of CO in the atmosphere of WASP-43b. Furthermore, we retrieved the temperature-pressure profile, abundances of CO and H$_2$O, and a super-solar C/O ratio of 0.78 by applying a Bayesian retrieval framework to the data. Our findings also shed light on the atmospheric characteristics of WASP-43b. We found no evidence for a cloud deck on the dayside, and recovered a line broadening indicative of an equatorial super-rotation corresponding to a jet with a wind speed of $\sim$ 5\,km\,s$^{-1}$, matching the results of previous forward models and low-resolution atmospheric retrievals for this planet.}


   \maketitle
%

\section{Introduction}

The study of hot Jupiters has opened up exciting opportunities and posed significant new challenges for the analysis of exoplanetary atmospheres. These gas giants provide an excellent opportunity for detailed observations of their atmospheres with current telescopes and instrumentation. Due to their large size and close proximity to the host star, they exhibit a relatively high planet-to-star flux ratio at near-infrared wavelengths, while the short orbital period allows for observations of full phase curves. Additionally, the high day-night temperature difference can cause fast winds on a global scale that redistribute heat and atmospheric constituents.

WASP-43b is a 2.0\,$M_\mathrm{Jup}$ hot Jupiter that was discovered by \citet{hellier2011WASP43bClosestorbitingHot}. The planet orbits a young K7 dwarf with a gyrochronological age estimate of 400$^{+200}_{-100}$\,Myr \citep{hellier2011WASP43bClosestorbitingHot}, which shows chromospheric and coronal emission consistent with a correspondingly high level of stellar activity \citep{czesla2013XrayIrradiationMassloss, staab2017SALTObservationsChromospheric}. Due to its short orbital period of only 0.81\,d and the small orbital distance of 0.015\,AU (see Table \ref{Table_Parameters}), WASP-43b is thought to be tidally locked, making it a suitable target to analyse variabilities across its atmosphere based on phase curve observations. Previous observational data include a low-resolution transmission spectrum and emission phase curves from the Hubble Wide Field Camera 3 \citep{bean2013FollowWaterUltimate}, as well as two datasets in the infrared region measured with the Spitzer InfraRed Array Camera \citep{blecic2014SPITZEROBSERVATIONSTHERMAL}. \citet{kreidberg2014PreciseWaterAbundance} first found evidence for H$_2$O absorption and emission, while \citet{chubb2020AluminiumOxideAtmosphere} recently reanalysed the data and reported evidence for the presence of AlO being favoured over all other investigated molecules. This points towards the existence of disequilibrium effects in the planet's atmosphere causing the heavy AlO molecules to be detectable in the upper regions. While the presence of other molecules, such as CO and CH$_4$, has been suggested in previous studies \citep{kreidberg2014PreciseWaterAbundance, feng2016ImpactNonUniformThermal, stevenson2017SpitzerPhaseCurve}, the results are not conclusive.

Despite WASP-43b's ultra-short period, its equilibrium temperature of $T_\mathrm{eq} \approx 1400\,$K is considerably lower than that of many other comparable planets due to the low temperature of its host star ($T_\mathrm{eff} \approx 4500\,$K). This opens up the possibility of cloud formation even on the dayside of WASP-43b, but the existence of such cloud layers is still under investigation. While \citet{helling2020MineralCloudHydrocarbon} and \citet{venot2020GlobalChemistryThermal} predict the presence of silicate and metal-oxide clouds based on 3D atmospheric modelling, \citet{scandariato2022PhaseCurveGeometric} analysed the geometric albedo and found no evidence for condensation of reflective clouds on the planetary dayside.

Exploring the C/O ratio of exoplanet atmospheres has been identified as a promising avenue towards unravelling the mysteries of planet formation and migration. \citet{oberg2011EffectsSnowlinesPlanetary} suggested that the C/O ratio could offer clues to the planet's origin within the protoplanetary disk, and several subsequent studies have explored this hypothesis \citep[e.g.][]{madhusudhan2016ExoplanetaryAtmospheresChemistry, mordasini2016ImprintExoplanetFormation, booth2019PlanetformingMaterialProtoplanetary, cridland2019ConnectingPlanetFormation}. The composition of solid and gaseous material varies with distance from the host star, owing to differences in their condensation temperatures, and as the planet forms and undergoes migration its final composition and atmospheric makeup are shaped by the materials it accumulates along the way. Using simulations, \citet{turrini2021TracingFormationHistory} find a significant difference in C/O based on whether the migration is gas dominated or solid enriched, while the dependence on the formation distance is less pronounced.

Whether the atmosphere is above or below the tipping point of C/O = 1 has major implications for the entire composition, especially for hot and ultra-hot Jupiters. If there is more carbon than oxygen present, the entire oxygen supply is bound in CO and is not available to form other oxides. For more moderate atmospheric temperatures, however, the dominant carbon-bearing species is CH$_4$, and thus the oxygen is free to form H$_2$O (e.g. \citealt{giacobbe2021FiveCarbonNitrogenbearing}).

Additionally, the observable C/O is linked to the formation of clouds, as the condensation of certain elements removes them from the gaseous phase that is accessible with transmission and emission spectroscopy. According to \citet{helling2021CloudPropertyTrends}, cloud particles such as MgO and Al$_2
$O$_3$ can remove significant amounts of oxygen from the local atmosphere, leading to an increase in C/O.

The previous observations of WASP-43b with the Hubble Space Telescope (HST) and Spitzer are suitable to measure the H$_2$O content in hot Jupiter atmospheres, but due to the limited wavelength coverage, the determination of other molecules with the majority of their strong emission lines in the infrared region is more difficult. This is the case for CO and CH$_4$, which emit most strongly at wavelengths longer than 2$\,\mu$m. However, an estimation of the C/O ratio necessarily relies on robust estimates of the abundances of these main carbon-bearing molecules. Studies of one and the same planet that are based on different observations, reduction methods, or model assumptions can lead to vastly different C/O measurements, as was the case for example for WASP-127b \citep[e.g.][]{spake2021AbundanceMeasurementsH2O, boucher2023ConoCo} or the ultra-hot Jupiter WASP-12b 
\citep[e.g.][]{madhusudhan2011HighRatioWeak, crossfield2012ReEvaluatingWASP12bStrong, stevenson2014DecipheringAtmosphericComposition}.

In the case of WASP-43b, the results are less controversial as they generally agree on a C/O < 1. \citet{kreidberg2014PreciseWaterAbundance} retrieved the abundances of the main carbon- and oxygen-bearing species using HST observations, but only the H$_2$O content was well constrained. Based on this result, they conclude that the C/O ratio is consistent with the solar value. \citet{benneke2015StrictUpperLimits} found an upper limit of C/O < 0.87 in a reanalysis of the data, while \citet{irwin20205DRetrievalAtmospheric} included the Spitzer/IRAC observations to retrieve C/O $\sim$ 0.91. However, they caution that the abundances of CO and CH$_4$ effectively only depend on the two Spitzer datasets. The data reduction and assumed thermal profile thus have a strong influence on the retrieved C/O ratio, such that \citet{changeat2021ExplorationModelDegeneracies} recovered C/O = 0.68 with slightly different methods and assumptions.

Using the large simultaneous wavelength coverage and high resolution (R $\sim$ 100\,000) of ground-based spectrographs such as CRIRES$^+$ mounted on the ESO VLT/UT3 telescope, it becomes possible to circumvent dependencies on a limited number of measurements and the ensuing ambiguities. Observing the spectral bands of CO and CH$_4$ in the mid-infrared promises the tightest constrains on their abundances due to the large number and intensity of spectral lines, consequently allowing for a more precise determination of the C/O ratio.

\begin{table}[ht]\renewcommand{\arraystretch}{1.5}
 \caption[]{Stellar and planetary parameters of the WASP-43 system.}\label{Table_Parameters}
\begin{tabular}{lll}
 \hline \hline
  Parameter &
  Symbol &
  Value

 \\ \hline
\textit{Planet}   &   & \\
Radius$^a$   & $R_p$ ($R_\mathrm{Jup}$) &  1.006 $\pm$ 0.017  \\
Mass$^a$   & $M_p$ ($M_\mathrm{Jup}$) &  1.998 $\pm$ 0.079  \\
Orbital period$^a$   & $P_\mathrm{orb}$ (days) & 0.813473978 \\
Orbital inclination$^a$    & $i$ ($^\circ$) &  82.109 $\pm$ 0.088\\
Orbital eccentricity$^a$   & $e$ & 0 \\
Semi-major axis$^a$   & $a$ (AU)& 0.01504 $\pm$ 0.00029\\
Time of mid-transit$^b$   & $T_0$ (BJD) & 2458555.80567 \\
RV semi-amplitude$^d$   & $K_p$ (km\,s$^{-1}$) & 202 $\pm$ 4 \\
Surface gravity$^d$   & $\log{g}$ (cgs)& 3.77 \\
Equil. temperature$^a$   & $T_\mathrm{eq}$ (K)& 1426.7 $\pm$ 8.5  \\
 \\ \hline
\textit{Star}   &   & \\
Radius$^a$   & $R_*$ ($R_\odot$) & 0.6506 $\pm$ 0.0054 \\
Effective temperature$^c$   & $T_\mathrm{eff}$ (K)& 4400 $\pm$ 200\\
Systemic velocity$^a$   & $v_\mathrm{sys}$ (km\,s$^{-1}$) & -3.5955 $\pm$ 0.0043 
 \\ \hline
\end{tabular}
\tablebib{
$^{(a)}$\citet{esposito2017GAPSProgrammeHARPSN},
$^{(b)}$\citet{patel2022EmpiricalLimbdarkeningCoefficients},
$^{(c)}$\citet{bonomo2017GAPSProgrammeHARPSN},
$^{(d)}$ Calculated from the other orbital parameters.
}
\end{table}

In this work, we present high-resolution dayside observations of WASP-43b by employing the CRIRES$^+$ (CRyogenic InfraRed Echelle Spectrograph $^+$, \citealt{dorn2014CRIRESExploringCold, dorn2023CRIRESSkyESO}) instrument. This spectrograph is the upgraded version of the original CRIRES \citep{kaeufl2004CRIRESHighresolutionInfrared}, offering a significantly larger simultaneous wavelength coverage and new infrared detectors. It is mounted on an 8\,m-class telescope, which allows for high S/N observations with sufficient time resolution. In contrast to previous observations using low-resolution instruments, the high resolution of CRIRES$^+$ offers the opportunity to directly measure individual absorption and emission lines, while the capability to observe in the infrared region enables the study of species such as CO and CH$_4$, which do not produce significant spectral features in the visible wavelength range.

In Sect. \ref{Section_Data} we describe the observations and our data reduction procedure, and in Sect. \ref{Section_method} we go into the details of the model generation, cross-correlation and retrieval analysis. We present our results and compare them to previous observations and simulations in Sect. \ref{Section_Results}, and conclude in Sect. \ref{Section_Conclusion}.

\section{Observation and data reduction} \label{Section_Data}

\begin{table}[ht]\renewcommand{\arraystretch}{1.5}
\caption{Observation log of the two nights. The variation of observational conditions over time is shown in Fig. \ref{Fig_NightlyConditions}.}             
\label{Table_ObservationLog}      
\centering          
\begin{tabular}{l|ll}
 \hline \hline
  Night &
  1 &
  2
 \\ \hline
    Date & 2022-03-11 & 2023-02-20\\
  Phase coverage & 0.52 - 0.66 & 0.28 - 0.49\\
  Exposure time per frame  & 300\,s & 300\,s\\
  N$_\mathrm{spectra}$ & 34 & 48\\
  Spectral resolution (R) & $\sim$\,120\,000 & $\sim$\,97\,000\\
  Mean S/N per exposure & 48 & 34\\
  Airmass  & 1.43 - 1.04 & 1.03 - 1.75\\
  \begin{tabular}{@{}c@{}}Median FWHM of the PSF \\ (in spatial direction)\end{tabular}  & 0.16 $\arcsec$& 0.33 $\arcsec$\\

\hline                  
\end{tabular}
\end{table}

\begin{figure}
\centering
\includegraphics[width=\hsize]{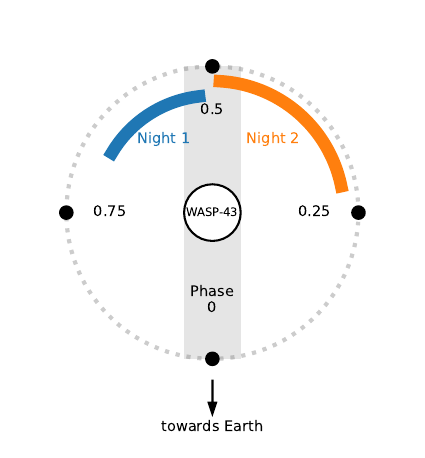}
  \caption{Orbital phase intervals of the two observations. The grey shaded area indicates the region in which the primary and secondary eclipses occur. The filled circles represent the planet WASP-43b, and the empty circle represents its host star. The semi-major axis and stellar radius are to scale.
          }
     \label{Fig_Phaseplot}
\end{figure}

We observed WASP-43b on 11 March 2022 and 20 February 2023 with the CRIRES$^+$ instrument as part of the guaranteed time observations of the CRIRES$^+$ consortium. 
The two observing runs were conducted shortly before and shortly after the secondary eclipse, respectively, to observe the dayside of WASP-43b (see Fig. \ref{Fig_Phaseplot} for the phase coverage). We used the nodding mode technique to observe the target alternating between two different positions (A and B) along the slit to allow for the removal of detector artefacts and the sky background during the extraction. In our further analysis, we combined the A and B spectra to form a single time series for each night. We observed in the K2192 wavelength setting covering parts of the wavelength region from 1972\,nm to 2486\,nm with an exposure time of 300\,s. The CRIRES$^+$ science detector array consists of 3 chips, and thus each spectral order is separated into three segments. In the analysis, we treated each segment individually, resulting in 20 wavelength segments in the chosen wavelength setting.

We employed metrology to improve the wavelength calibration by fine-tuning the position of a defined set of emission lines from Kr- and Ne-lamps, and the 0.2$\arcsec$ slit was chosen to ensure a high spectral resolution. The first observing run was conducted under good weather with stable seeing and relative humidity ($\sim$30\%), and adaptive optics (AO) was employed. The second night was hampered by a varying coverage of thick and thin clouds, especially during the first half of the observation, which prohibited the use of AO and caused a substantial loss of flux. Although the seeing was similar in both nights, this led to a vastly increased point spread function (PSF) during the second night. The observational conditions that could influence the data quality are shown in Fig. \ref{Fig_NightlyConditions} for the two nights.

The raw spectra were reduced with the ESO CRIRES$^+$ data reduction pipeline (version 1.2.3) using the \texttt{EsoRex}\footnote{https://www.eso.org/sci/software/cpl/esorex.html} tool.  This pipeline includes the standard reduction steps for echelle time series observations, such as dark, bias and flat field correction, removal of bad pixels, and a calculation of the wavelength solution. In addition, we attempted to refine the wavelength solution by using Molecfit \citep{smette2015MolecfitGeneralTool} to fit the telluric lines in the spectra, but we did not find any significant offset to the solution provided by the pipeline. In addition, by cross-correlating the telluric lines between individual spectra we measured the drift of the gratings, which manifested themselves in a slight shift of the spectra in time. The calculated drifts were consistently smaller than 0.1 pixels (corresponding to a velocity shift of $\sim$$ 0.1\,$km\,s$^{-1}$), and therefore we did not apply any further correction.

In the case of good observing conditions that allow an excellent AO correction and coherence time, the PSF can be smaller than the minimum slit width of 0.2 $\arcsec$, causing an increase in the spectral resolution \citep{dorn2023CRIRESSkyESO}. The median FWHM of the PSF during night 1 was 2.8 pixel (corresponding to \hbox{0.16 $\arcsec$}), while in night 2 it was 5.8 pixel (0.33 $\arcsec$).
To determine the resolution, we assumed that the width of the PSF in dispersion direction is similar to the one in spatial direction. Then we converted the PSF from pixel space into wavelength space, which allows for the calculation of the resolution for each segment, and found an average resolution of R $\sim$\,120\,000 for night 1. The PSF in night 2 was significantly larger, and in this case the resolution was instead constrained by the slit width to R $\sim$\,97\,000. This shows the effect of the cloud coverage and consequent lack of AO in the second observation, resulting in a significantly broader PSF. Details of the two observations are summarised in Table \ref{Table_ObservationLog}.

\subsection{Normalisation}
In the following, we describe the normalisation of the spectra, which consists of removing outliers, masking any sky emission lines, and removing the continuum variations.
For the outlier removal, any data points deviating by more than 5$\sigma$ in the time series of each pixel were removed and linearly interpolated. If more than three values of a pixel's time series would have to be replaced, the entire pixel was masked for all spectra of the time series. 

To mask the sky emission lines, a rough continuum approximation was computed and all regions with a flux above 103\% of the continuum were masked. To compute such a continuum for each segment, a 90\% percentile filter with a width of 100 pixel was applied to the spectra, before smoothing the result with a 50 pixel wide Gaussian filter.

To remove flux variations in the continuum, we determined some points following the shape of the continuum, to which we can fit a polynomial. To this end, we computed a master spectrum for each order as the time average of all exposures. We divided the master spectrum into 100 wavelength bins of equal size, and assigned to each bin the 90\% percentile value of all flux values in the bin. Subsequently, the 100 points found in this way were binned again into only ten bins, and from each of these bins we selected the second largest value. The ten resulting points follow the overall shape of the blaze function while not being skewed by improperly removed emission lines. A third degree polynomial was fitted to these ten data points, and all spectra and errors were divided by the fit. 
Subsequently, each individual spectrum of each segment was corrected to the same continuum level by division of a linear fit to the entire spectrum. As the last normalisation step, deep telluric lines that drop below 40\% of the continuum flux were masked.

\subsection{SYSREM}
We used SYSREM (\citealt{tamuz2005CorrectingSystematicEffects, birkby2013DetectionWaterAbsorption}, see also \citealt{nugroho2017HighresolutionSpectroscopicDetection, alonso-floriano2019MultipleWaterBand, cabot2019RobustnessAnalysisTechniques, gibson2020DetectionFeAtmosphere}) for the removal of the telluric and stellar lines from the spectra. This principal component analysis algorithm takes into account the uncertainty of each data point to construct a model of the data consisting of linear components in wavelength and time. The stellar, telluric and planetary components of the observed spectra all have different Doppler shifts. While the stellar and telluric lines can be considered nearly static in velocity space over the course of the observation, the planetary radial velocity changes significantly  of the order of several tens of km\,s$^{-1}$. The SYSREM algorithm iteratively subtracts linear trends in time from each spectral pixel, which primarily serves to remove the (quasi-)static components of the spectra. This leaves residuals consisting only of noise and the planetary signal, which shifts in wavelength over time due to the planetary motion and is therefore not modelled by SYSREM. During each iteration, the linear model created by SYSREM is continuously refined until a convergence criterion is met, after which the model is subtracted from the data and the computation of the next iteration starts. We assumed the model of each iteration to be converged when the average relative change is below 0.01.

\begin{figure}
\centering
\includegraphics[width=\hsize]{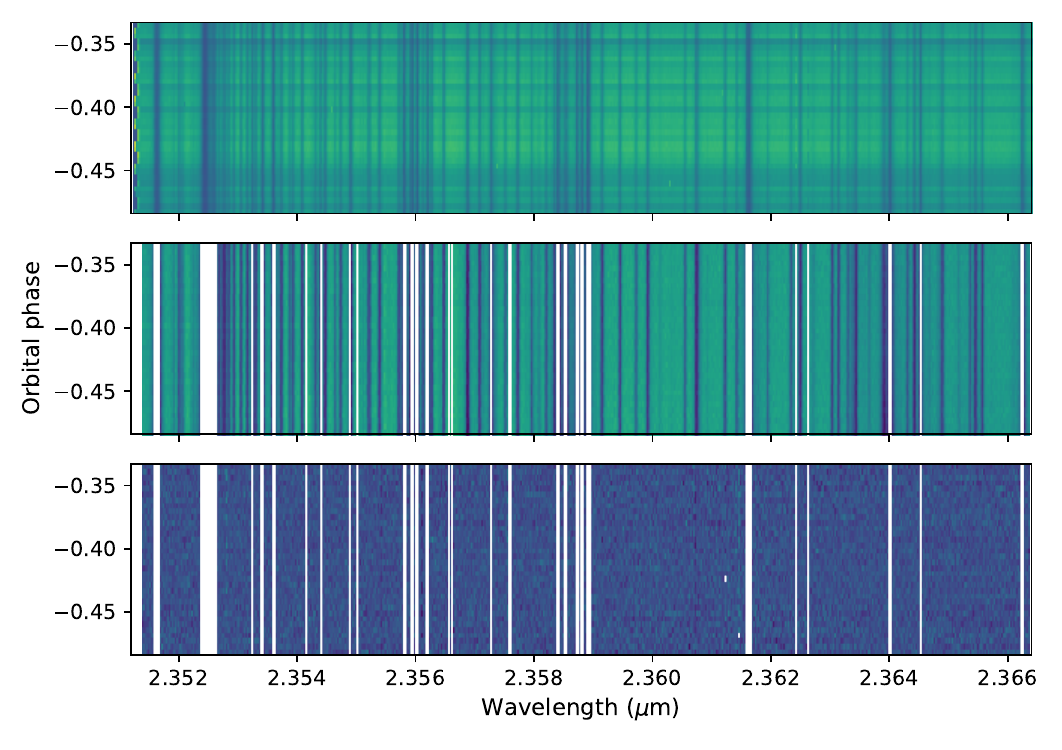}
  \caption{Data reduction steps of a representative wavelength range. \textit{Top panel:} Unprocessed spectra as they are produced by the CRIRES$^+$ pipeline. \textit{Middle panel:} After continuum normalisation and masking of deep telluric lines. \textit{Bottom panel:} After the removal of stellar and telluric lines with SYSREM.
          }
     \label{Fig_DataReduction}
\end{figure}

 The number of SYSREM iterations applied to the data can have a large influence on the recovered signal strength, and we explain our process for choosing this number in Sect. \ref{Sysrem iterations}.
The effect of the normalisation and application of SYSREM on the spectra is illustrated in Fig. \ref{Fig_DataReduction}.

\section{Methods} \label{Section_method}
Our analysis consisted of generating synthetic planet spectra with key molecular species of interest, cross-correlating these with our cleaned CRIRES$^+$ spectra and subsequently applying a retrieval framework to constrain the atmospheric properties.

\subsection{Generating planetary model spectra} \label{Model generation}

The synthetic model spectra for the cross-correlation analysis were computed with the radiative transfer code petitRADTRANS \citep{molliere2019PetitRADTRANSPythonRadiative}, using the CO, H$_2$O and CO$_2$ line lists of \citet{rothman2010HITEMPHightemperatureMolecular}, the CH$_4$ line list of \citet{hargreaves2020AccurateExtensivePractical}, the HCN line list of \citet{harris2006ImprovedHCNHNC} and \citet{barber2014ExoMolLineLists}, and the NH$_3$ line list of \citet{yurchenko2011VariationallyComputedLine}. We assumed a non-inverted temperature-pressure ($T$-$p$) profile characterised by the parametric model from \citet{guillot2010RadiativeEquilibriumIrradiated}. This model is determined by the irradiation temperature $T_\mathrm{irr}$, the internal temperature $T_\mathrm{int}$, the infrared opacity $\kappa_\mathrm{IR}$ and the ratio of visible to infrared opacity $\gamma$ (see his Eq. 29). 
Based on the assumed volume mixing ratios (VMRs), we calculated the mean molecular weight and subsequently the mass fractions of these molecules, assuming a solar-like VMR ratio between H$_2$ and He.
We included collision-induced absorption (CIA) of H$_2$-H$_2$ and H$_2$-He \citep[and the references therein]{borysow2002CollisioninducedAbsorptionCoefficients, richard2012NewSectionHITRAN}, which we found to have a substantial effect on the line strength for this planet. For analysing the influence of clouds, we additionally added a grey cloud deck at varying pressures, which blocks the contribution of deeper layers. The resulting spectra were converted to a planet-to-star flux ratio.
In addition we accounted for rotational broadening of the lines, which has a non-negligible effect for high-resolution spectroscopy. Following Eq. (3) in \citet{diaz2011AccurateStellarRotational}, a rotational profile was computed based on the equatorial rotation velocity $v_\mathrm{eq}$. Similar to \citet{yan2023CRIRESDetectionCO}, we assumed a linear limb darkening law with a coefficient $\epsilon = 1$ and an inclination angle of $\sin{i} \approx 1$ as the planet is expected to be tidally locked and thus the equatorial plane is aligned with the orbital plane. The model was then convolved with this profile in logarithmic wavelength space.
Subsequently, the spectra were convolved to the previously determined resolution of CRIRES$^+$ (Sect. \ref{Section_Data}) using a Gaussian instrumental profile, and continuum normalised.

\begin{figure}
\centering
\includegraphics[width=\hsize]{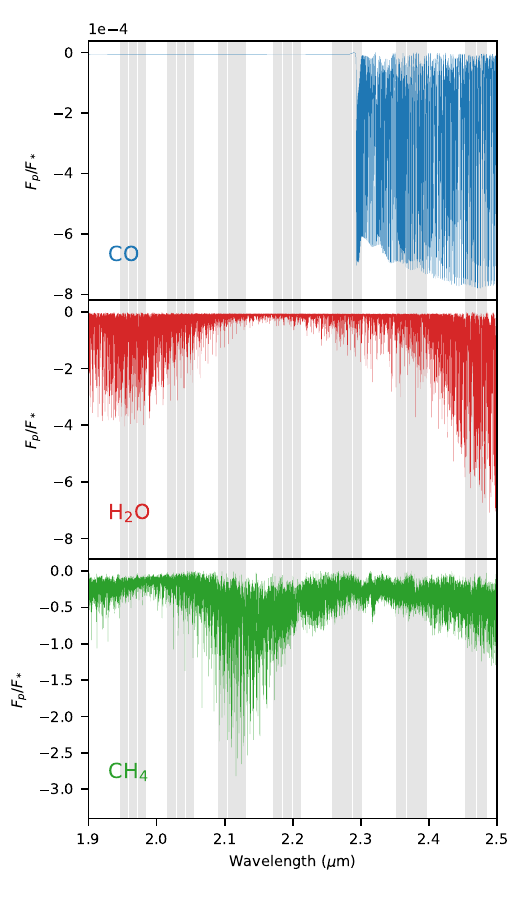}
  \caption{Synthetic model spectra of CO (\textit{top panel}), H$_2$O (\textit{middle panel}) and CH$_4$ (\textit{bottom panel}) in the wavelength range of the observations, calculated using petitRADTRANS. The shown models are convolved to a resolution of  R = 100\,000 and are not rotationally broadened. The grey regions indicate the wavelength ranges covered by the K2192 setting.
          }
     \label{Fig_Models}
\end{figure}

We computed individual synthetic spectra for the species analysed with cross-correlation, and incorporated the model calculation in our retrieval framework. Figure \ref{Fig_Models} shows the CO, H$_2$O and CH$_4$ models used for cross-correlation, before rotational broadening is applied.

\subsection{Cross-correlation} \label{Section_CC}
	
\begin{figure}
\centering
\includegraphics[width=\hsize]{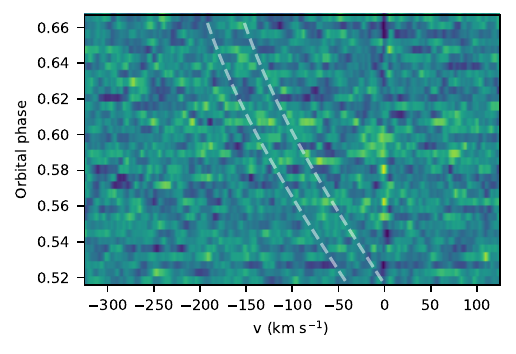}
  \caption{Cross-correlation function of CO for night 1 after five sysrem iterations. The dashed lines indicate the expected planetary trail after the eclipse (which is located between the two lines). The noise structure at $\varv = 0 $ km$s^{-1}$ is caused by tellurics that could not be completely removed.
          }
     \label{Fig_CCF_CO_Night1}
\end{figure}

We calculated the weighted cross-correlation function (CCF) from the residual spectra and the model following the work from \citet{cont2022AtmosphericCharacterizationUltrahot}:
\begin{align}
    \mathrm{CCF}(v, t) = \sum _{i = 0}^N \frac{R_i(t) \cdot M_i(v)}{\sigma_{i}(t)^2}\,,
\end{align}
where $t$ is the time index, $i$ is the pixel index, $R$ are the residual spectra, $M$ is the model spectrum and $\sigma$ is the uncertainty of $R$. The model spectrum was Doppler-shifted with velocities $v$ ranging from -500\,km\,s$^{-1}$ to +500\,km\,s$^{-1}$. This created a two-dimensional CCF map for each spectral segment. Combining the information of all segments into a single map was done by computing the mean CCF map. In this step we excluded some segments that were substantially contaminated by tellurics and in which no planetary signal could be recovered. We selected these segments by first visually inspecting the spectra, and subsequently calculating the signal strength of an injected model of CO and H$_2$O as outlined in Sect. \ref{Sysrem iterations} to determine whether the signal improves when a certain segment is excluded. Based on these criteria, we excluded segments 1 and 20 in night 1 and segments 1, 2 and 20 in night 2. An example of the CCF map for night 1 is shown in Fig.  \ref{Fig_CCF_CO_Night1}.

Following this, a $K_p$-$v_\mathrm{sys}$ map was calculated from the CCFs. To this end, we considered a range of orbital semi-amplitude values ($K_p$), and for each value we shifted the CCFs into the corresponding planetary frame according to the Doppler velocity:
\begin{align}
    v_p = K_p \sin\,(2\pi \,\phi) + v_\mathrm{sys} - v_\mathrm{bary} + v_\mathrm{max}\,, \label{Eq_velocity}
\end{align}
where $\phi$ is the orbital phase, $v_\mathrm{sys}$ and $v_\mathrm{bary}$ are the systemic and barycentric velocities and $v_\mathrm{max}$ is a potential deviation from the planetary rest frame. Subsequently, the shifted CCFs were collapsed into a one-dimensional vector by averaging along the time axis. Stacking these vectors for every trial $K_p$ value resulted in a two-dimensional $K_p$-$v_\mathrm{sys}$ map, which was converted into units of S/N by dividing by the standard deviation of the map in regions far away from the expected signal position ($|v| > 50$\,km\,s$^{-1}$).

\subsection{Selecting the number of SYSREM iterations} \label{Sysrem iterations}
\begin{figure}
\centering
\includegraphics[width=\hsize]{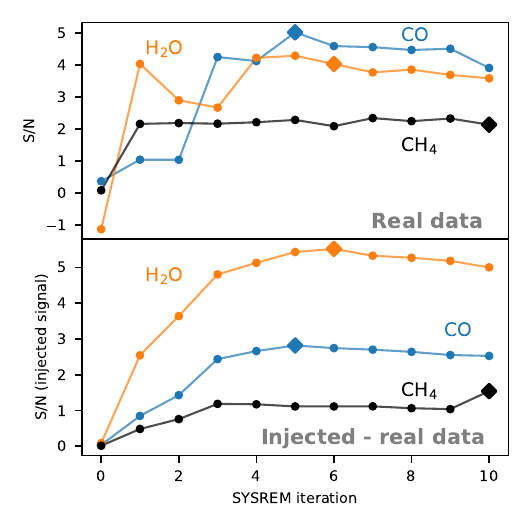}
  \caption{S/N detection strength as a function of SYSREM iterations for the data of night 1. The measured signal strengths are a result of using a pure CO model (blue colour), a pure H$_2$O model (orange colour), and a pure CH$_4$ model (black colour). The \textit{top panel} shows the signal strength for the real data, while the \textit{bottom panel} shows the strength of an injected signal using the differential $\Delta$CCF according to the method described in Sect. \ref{Sysrem iterations}. The diamond shape indicates the SYSREM iteration with the strongest detection of the injected signal.
          }
     \label{Fig_SNR-Sysrem_iteration}
\end{figure}

The removal of stellar and telluric lines with SYSREM is an iterative process, where the choice of the number of iterations influences the final result. Too few iterations lead to strong residuals in the $K_p$-$v_\mathrm{sys}$ map which may hinder the detection of the planetary spectral signature. With too many iterations, SYSREM begins to affect and remove the planetary trail as well. In order to identify the optimal number of iterations in an objective way, we followed the method proposed by \hbox{\citet[see their Sect. 3.8]{cheverall2023RobustnessMeasuresMolecular}}. We first applied SYSREM to the normalised data and calculated the cross-correlation function CCF$_\mathrm{obs}$. Then we repeated these analysis steps on the same data, but with a planetary model spectrum injected at the expected Doppler velocity. This yielded the signal-injected cross-correlation function CCF$_\mathrm{inj}$. Subsequently, we derived the differential cross-correlation function $\Delta$CCF = CCF$_\mathrm{inj}$ - CCF$_\mathrm{obs}$, and calculated the $K_p$-$v_\mathrm{sys}$ map and the S/N for each iteration. Finally, we chose the SYSREM iteration that maximises the S/N of $\Delta$CCF, and applied the same number of iterations to the non-injected data. The results of this selection process are summarised in Table \ref{Table_CCresults} and Fig. \ref{Fig_SNR-Sysrem_iteration}.

\subsection{Retrieval framework} \label{Section Retrieval framework}

The properties of the atmosphere of WASP-43b were analysed by performing an atmospheric retrieval, in which the observed data is compared to a large number of different models. Using a sampling algorithm to determine the posterior distributions of the model parameters allows for the statistical measurement of atmospheric characteristics. In this way, any properties that have an effect on the observed emission spectra can be analysed. Both the $T$-$p$ profile and the abundances influence the line strengths, while the position of the lines over the course of the observing run depends on the $K_p$ and systemic velocity. Additionally, the equatorial velocity $v_\mathrm{eq}$ can be determined as a rotation increases the line width by means of rotational broadening. The existence of clouds at a certain pressure P$_\mathrm{cloud}$ truncates the spectral lines across the entire wavelength range as contributions from regions below the cloud deck are blocked.

For the retrieval, we assumed that the abundances are isobaric and that the temperature is described along the parameterisation of \citet{guillot2010RadiativeEquilibriumIrradiated}. While this profile does not have enough degrees of freedom to capture the temperature variations across the entire height of the atmosphere, we expect it to be sufficient to model the region with the steepest temperature gradient, from which the majority of the signal originates. We did not include Rayleigh scattering as this process did not have any appreciable effect on the models.

We followed the method of \citet{yan2023CRIRESDetectionCO}, which is based on the work of \citet{yan2020TemperatureInversionAtomic} and inspired by \citet{brogi2019RetrievingTemperaturesAbundances}, \citet{shulyak2019RemoteSensingExoplanetary} and \citet{gibson2020DetectionFeAtmosphere}. The forward models were computed as described in Sect. \ref{Model generation}. To be able to compare these models to the data, it is important to consider any factors that impact the planetary signal. Therefore we modelled the effect of smeared spectral lines and applied corrections to account for distortions introduced by the application of SYSREM.

\subsubsection{Accounting for smearing of the signal} \label{Section_Smearing}
\begin{figure}
\centering
\includegraphics[width=\hsize]{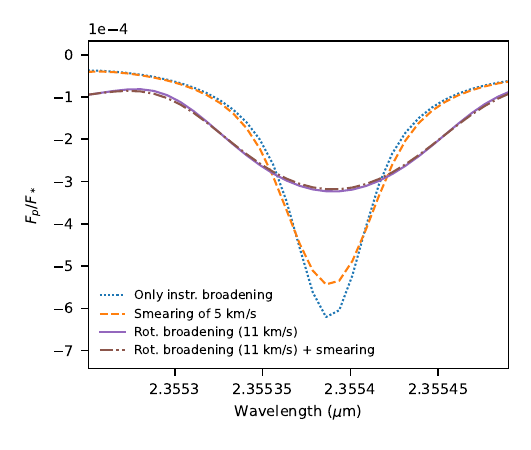}
  \caption{Profile of a single CO line for models with different combinations of smearing and rotational broadening. All models have the instrumental broadening applied, and were additionally smeared with a velocity change of 5\,km\,s$^{-1}$ (orange), rotationally broadened with $v_\mathrm{eq} = 11$\,km\,s$^{-1}$ (purple), or had both effects combined (brown).
          }
     \label{Fig_BroadeningEffects}
\end{figure}

Due to its short orbital period, the radial velocity of WASP-43b changes significantly during a single exposure of 300\,s. For night 1, the velocity difference between start and end of an exposure was between 2 - 5 km\,s$^{-1}$, causing a substantial smearing of the spectral lines. To account for this effect, the model was interpolated to the same wavelength grid as the 2D data (wavelength $\times$ time), resulting in a model matrix $M$. The model could then be shifted according to the planetary Doppler velocity 
(see Eq. \ref{Eq_velocity}). For each exposure, we calculated these radial velocities at the beginning and end time, shifted the model to ten evenly spaced planetary rest-frames between these two velocities, and computed the smeared model as the mean over the ten sub-exposures. The effect of smearing on the shape of the spectral lines is shown in Fig.  \ref{Fig_BroadeningEffects}. For a model without rotational broadening, the smearing changes the line shape significantly, while in the case of a strongly rotationally broadened model, the smearing is barely noticeable.

\subsubsection{Filtering the models}
Applying SYSREM not only removes static signals in the data, but also distorts the remaining planetary signal. Therefore, it is important to account for this distortion effect when comparing the planetary signal with synthetic model spectra via Bayesian inference algorithms to retrieve precise atmospheric properties. One could apply the same detrending procedure to both data and model, but this is computationally expensive. Instead, \mbox{\citet{gibson2022RelativeAbundanceConstraints}} proposed a fast alternative filtering technique using just the output of SYSREM, which is briefly summarised in the following:

During each iteration, SYSREM calculates column and row vectors \textbf{u} and \textbf{w} as a best approximation of the data. Combining all column vectors into a matrix \textbf{U} (with the dimensions $N_\mathrm{exposures}$ $\times$ $N_\mathrm{iterations}$), the correction for a model $M_\mathrm{unfiltered}$ can be computed:

\begin{align}
    M' = U(\Lambda U)^\dagger(\Lambda M_\mathrm{unfiltered})\,,
\end{align}
where $A^\dagger = (A^TA)^{-1}A^T$ is the Moore-Penrose inverse of a matrix $A$. Each entry in the diagonal matrix $\Lambda$ represents the reciprocal mean uncertainty in wavelength direction of the corresponding exposure. We precomputed this filter and applied it to every model during the retrieval, yielding the filtered model $M_\mathrm{unfiltered} - M'$ which was then compared to the data.

\subsubsection{Likelihood function}
To retrieve the atmospheric properties, we applied a Markov chain Monte Carlo (MCMC) algorithm using the $\texttt{EMCEE}$ package \citep{foreman-mackey2013EmceeMCMCHammer} to the logarithm of the likelihood function \citep{hogg2010DataAnalysisRecipes, yan2020TemperatureInversionAtomic}:

\begin{align}
\ln(L) = - \frac{1}{2}\sum_{i,j}\left(\frac{(R_{i,j} - M_{i,j})^2}{(\beta \sigma_{i,j})^2} + \ln(2\pi(\beta \sigma_{i,j})^2)\right)\,,
\end{align}
where $R_{i,j}$ is the residual spectrum at pixel index $i$ and time $j$, $M_{i,j}$ is the 2D matrix of a filtered model spectrum, $\sigma_{i,j}$ are the uncertainties of the residuals and $\beta$ is a scaling factor to $\sigma_{i,j}$. For the MCMC, we used 32 walkers with 15\,000 steps each. We chose uniform priors for all input parameters, which are summarised in Table \ref{Table_Priors}.

\section{Results} \label{Section_Results}

\subsection{Cross-correlation results}

\begin{figure}
\centering
\includegraphics[width=\hsize]{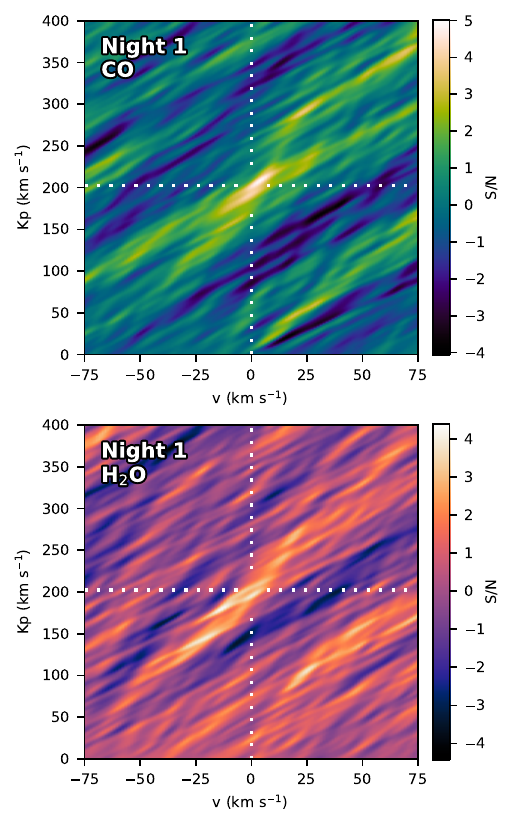}
  \caption{$K_p$-$v_\mathrm{sys}$ map for the CO signal after five SYSREM iterations (\textit{top panel}) and H$_2$O signal after six iterations (\textit{bottom panel}) in night 1. The white dotted lines indicate the expected position of a signal at the literature $K_p$ value in the planetary rest-frame. The CO signal has a S/N of 5.02, and the H$_2$O signal a S/N of 4.03.
          }
     \label{Fig_DetMap_night 1}
\end{figure}

We calculated $K_p$-$v_\mathrm{sys}$ maps for CO, H$_2$O and CH$_4$ as detailed in Sect. \ref{Section_CC}. The models used for the cross-correlation were computed assuming abundances of $\log_{10}$ CO = $-4$, $\log_{10}$ H$_2$O = $-4$ and $\log_{10}$ CH$_4$ = $-4$, respectively. We calculated the temperature-pressure profile using the parameterisation of \citet{guillot2010RadiativeEquilibriumIrradiated} with $T_\mathrm{irr} = 1400\,$K, $T_\mathrm{int} = 800\,$K, log$_{10}(\kappa_\mathrm{ir})$= $-2$ and log$_{10}(\gamma) = -0.4$, and calculated individual model spectra for these species. We only included instrumental broadening and did not account for any other broadening effects for these models. The cross-correlation analysis was conducted separately for each night. Table \ref{Table_CCresults} summarises the resulting S/N of the signal in the $K_p$-$v_\mathrm{sys}$ maps, as well as the number of SYSREM iterations, determined as outlined in Sect. \ref{Sysrem iterations}. Figures \ref{Fig_DetMap_night 1} and \ref{Fig_DetMap_Night2} show the $K_p$-$v_\mathrm{sys}$ maps for CO and H$_2$O of night 1 and night 2, respectively, while the results for CH$_4$ are shown in Fig. \ref{Fig_DetMap_CH4}.

\begin{table}[ht]\renewcommand{\arraystretch}{1.5}
 \caption[]{S/N of the signal in the $K_p$-$v_\mathrm{sys}$ map.}\label{Table_CCresults}
\begin{tabular}{l|l|ccc}
 \hline \hline
 &
  Night &
  CO &
  H$_2$O &
  CH$_4$
 \\ \hline
S/N & 1 &  5.02 &  4.03 & 2.14\\
& 2 & 0.82  & 2.47 & 2.46\\
\hline
\# iterations & 1 & 5 & 6 & 10\\
& 2 & 8 & 8 & 8\\
\end{tabular}
\end{table}

We detected the presence of CO and evidence for H$_2$O in the first night. However, we were not able to find either of the species in night 2. This does not come as a surprise, because the data quality of night 2 is considerably worse (see Sect. \ref{Section_Data}). We assessed the quality of night 2 by injecting a signal with the expected strength, and were not able to recover it. Therefore the non-detection in night 2 is most likely due to the poor data quality. \citet{pino2022GAPSProgrammeTNG} showed that principal component analysis algorithms remove much of the planetary signal at phases close to quadrature, as there is only a small shift in the line positions. This could affect especially the signal in night 2, as the observation started at a phase of 0.28. Using an injection-recovery test, we determined that SYSREM removes a substantial amount of the planet signal at phases up to 0.4. We ran the analysis again and excluded any exposures prior to this point in orbit, but we were still not able to detect CO or H$_2$O in night 2.

We found no evidence for CH$_4$ in either of the nights. In addition we searched for other molecules that are expected to be present based on equilibrium chemistry calculations (CO$_2$, HCN and NH$_3$), but we found no evidence for any of these species in either night.
We note that there are spurious peaks which have a S/N of 3-4 visible in some of the $K_p$-$v_\mathrm{sys}$ maps. However, these are far away from the expected position of a real signal, and are generally not stable against small changes in the model. We tested different models with variations in the $T$-$p$ profile or the inclusion of rotational broadening, and found that the strength and position of these spurious peaks is not consistent. On the other hand, the CO and H$_2$O signal in night 1 remain at the same position and at a stable S/N for all of the tested models. This shows that one has to be careful in the interpretation of signals with S/N $\sim$ 4, as these can in principle be created by spurious noise structures, and the stability of such signals against changes in the model should be investigated.

The results of night 1 agree with earlier studies of WASP-43b using HST and Spitzer observations. Since the first detection of H$_2$O by \citet{kreidberg2014PreciseWaterAbundance}, water features have been continuously reconfirmed in subsequent analyses \citep{stevenson2017SpitzerPhaseCurve, chubb2020AluminiumOxideAtmosphere}. While these studies found no conclusive evidence for CO absorption or emission lines, this is not surprising as the K band, which hosts the most prominent spectral bands of CO, was not covered by these previous observations.

\begin{figure*}[h]
  \includegraphics[width=\textwidth]{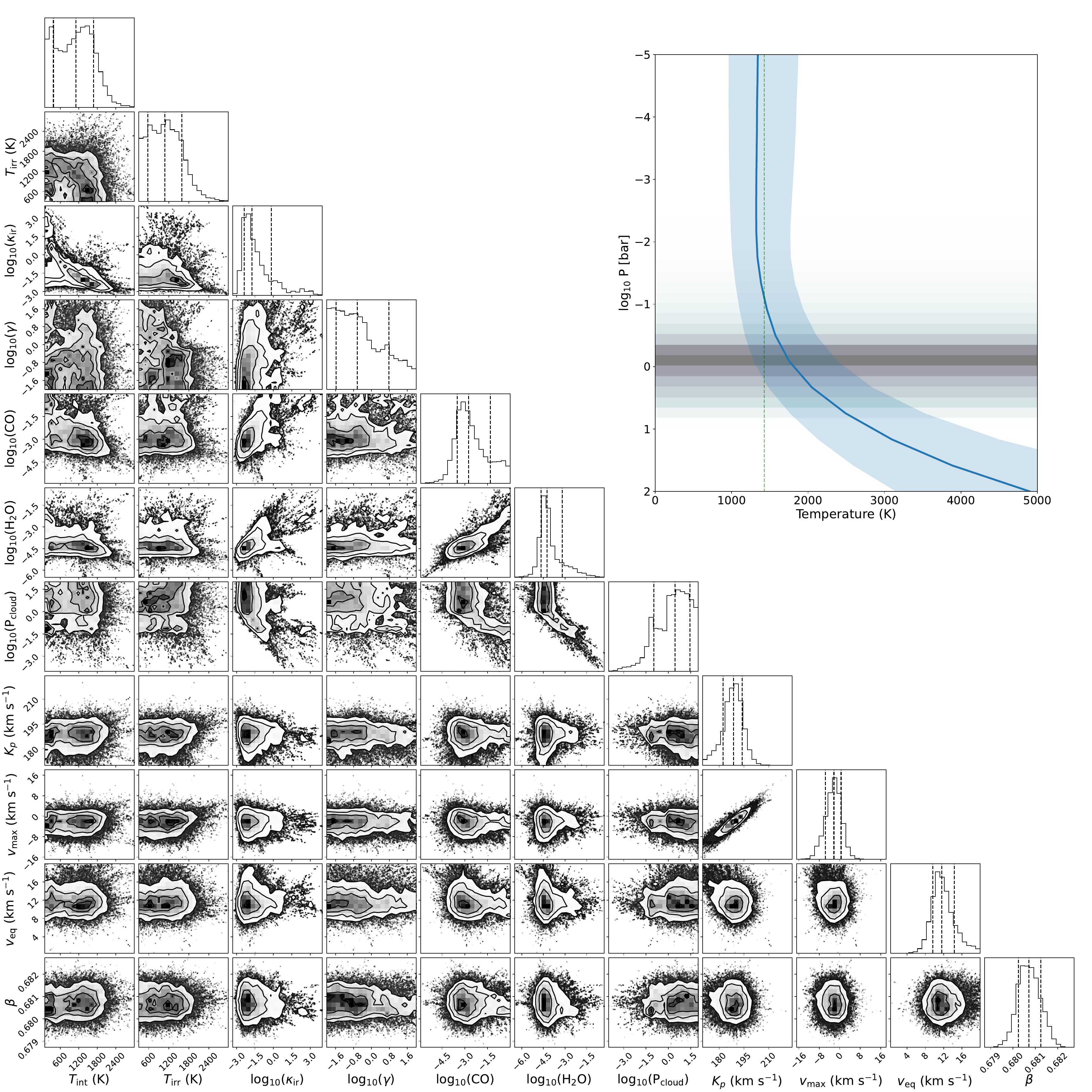}
  \caption{Posterior distributions of the atmospheric and orbital parameters for WASP-43b using the data of night 1. The top right panel shows the median $T$-$p$ profile, which was computed from 10\,000 random samples of the MCMC walkers. The blue shaded region indicates the 1$\sigma$ uncertainty interval, while the vertical dotted line shows the equilibrium temperature of WASP-43b ($T_\mathrm{eq} = 1426.7\,$K). The mean contribution function for the wavelength range of our observation is shown in grey.}
  \label{Fig_Cornerplot}
\end{figure*}

\subsection{Retrieval results}

\begin{table}[ht]\renewcommand{\arraystretch}{1.5}
 \caption[]{Priors and retrieved values of the retrieval using only the data from night 1.}\label{Table_Priors}
\begin{tabular}{llll}
 \hline \hline
  Parameter &
  Prior &
  Retrieved value &
  Unit

 \\ \hline
$T_\mathrm{int}$   & [100, 3000] & 1100$^{+600}_{-700}$ &  K  \\
$T_\mathrm{irr}$   & [300, 3000] & 1100 $\pm$ 500 &  K \\
$\log_{10}$ $\kappa_\mathrm{IR}$   & [-15, 4] & $-1.7^{+1.4}_{-0.6}$ &  $\log_{10}$ cm$^2$\,s$^{-1}$  \\
$\log_{10}$ $\gamma$  & [-2, 2] & $-0.6^{+1.4}_{-1.0}$  &   ... \\
$\log_{10}$ CO  & [-10, 0] & $-2.8^{+1.4}_{-0.8}$  &  ...  \\
$\log_{10}$ H$_2$O   & [-10, 0] & $-4.3^{+1.0}_{-0.4}$  &   ... \\
$\log_{10}$ P$_\mathrm{cloud}$ & [-8, 2] & $0.5^{+1.0}_{-1.4}$  &  $\log_{10}$ bar  \\
$K_p$   & [170, 230] & $188^{+5}_{-6}$  &  km\,s$^{-1}$  \\
$v_\mathrm{max}$   & [-20, 20] & $-2\pm 3$  &  km\,s$^{-1}$  \\
$v_\mathrm{eq}$   & [0.1, 20] & $11.7^{+2.8}_{-2.1}$  &  km\,s$^{-1}$  \\
$\beta$   & [0.2, 5] & $0.6807\pm 0.0005$  &   ... \\
\hline
\end{tabular}

\end{table}

While the cross-correlation method can confirm the presence of a species, it does not provide robust estimates of the abundances. We therefore applied the retrieval framework detailed in Sect. \ref{Section Retrieval framework} to constrain the atmospheric properties. As we did not find evidence for a signal of CO or H$_2$O (or any of the other species) in night 2, we applied the retrieval only to the data of night 1.

Following a 15\,000 step MCMC sampling, we discarded the initial 5000 steps as burn-in, and subsequently calculated the posterior distributions of the atmospheric and orbital parameters, which are illustrated in Fig.  \ref{Fig_Cornerplot}. The retrieved parameters and their uncertainties are given in Table \ref{Table_Priors}. We retrieved a non-inverted $T$-$p$ profile (see Fig. \ref{Fig_Cornerplot}), which agrees in the upper atmospheric layers with the equilibrium temperature of $T_\mathrm{eq} = 1426.7 \pm 8.5 \,$K. We found a planet velocity semi-amplitude $K_p = 188^{+5}_{-6}\,$km\,s$^{-1}$ and an offset in velocity from the planetary rest frame of $v_\mathrm{max} = -2\pm 3\,$km\,s$^{-1}$. The retrieved $K_p$ value is slightly below the literature value of $K_{p, \mathrm{lit}} = 202 \pm 4\,$km\,s$^{-1}$, which can be explained by the elongated diagonal structure of the H$_2$O detection peak in the $K_p$-$v_\mathrm{sys}$ map and the dependency of $K_{p, \mathrm{lit}}$ on the semi-major axis, which is not precisely known.

\begin{figure*}
  \includegraphics[width=\textwidth]{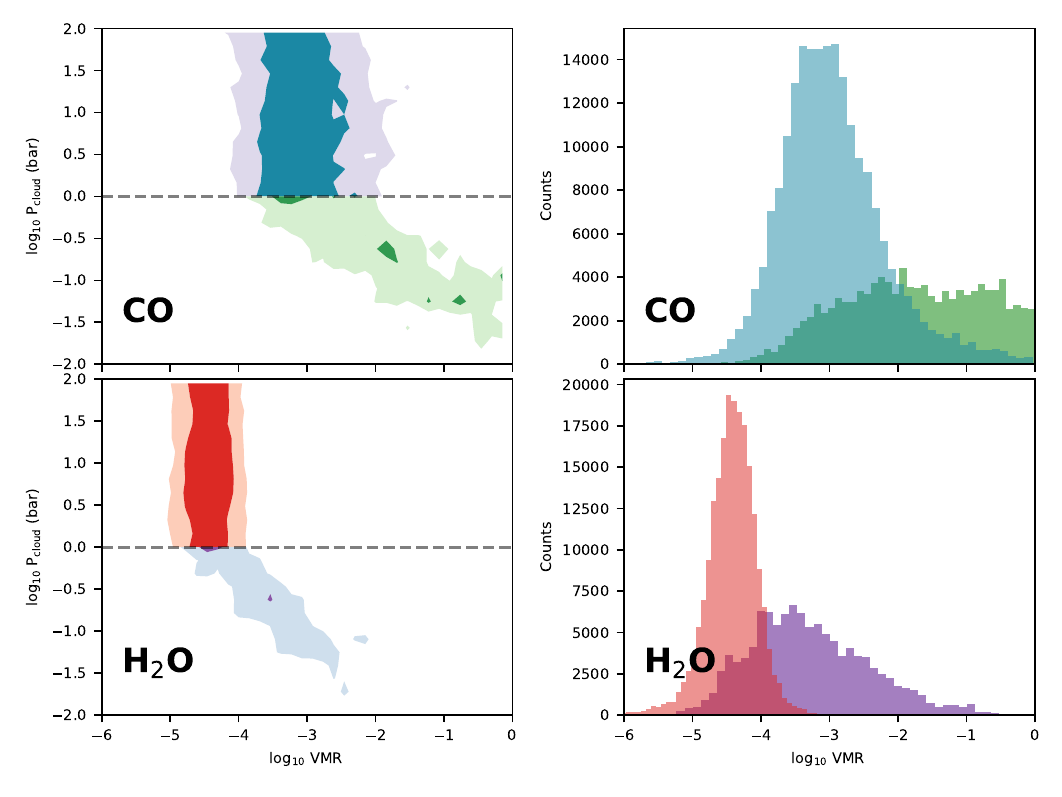}
  \caption{Posterior distributions of the CO abundance (\textit{top panels}) and H$_2$O abundance (\textit{bottom panels}). The \textit{left panels} show the correlation between the VMRs and the cloud deck pressure, and the  \textit{right panels} show the total distribution of the VMRs divided into two subsets: with $\log_{10}$ P$_\mathrm{cloud}$ > 0 (blue and red) and with $\log_{10}$ P$_\mathrm{cloud}$ < 0 (green and purple). The threshold (indicated by the dashed line) is caused by collision-induced absorption, which blocks any signals originating in the deeper layers.}
  \label{Fig_Cloud-Abundances}
\end{figure*}

\begin{figure}
\centering
\includegraphics[width=\hsize]{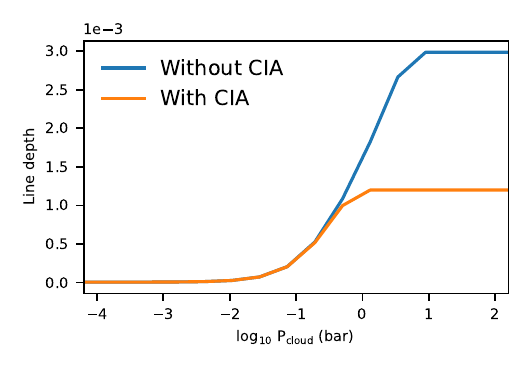}
  \caption{Line depth of the deepest line in a model with CO and H$_2$O as a function of cloud deck pressure, for a model with collision-induced absorption (orange) and without this effect (blue).
          }
     \label{Fig_CIA_influence}
\end{figure}

The inclusion of a cloud deck leads to a muting of the signal, as contributions of layers below the cloud deck pressure $P_\mathrm{cloud}$ become inaccessible to observations. This introduces a degeneracy between the cloud height and abundances, as a muting of the signal due to clouds can be counterbalanced by increased volume mixing ratios. Figure \ref{Fig_Cloud-Abundances} shows the correlation between $P_\mathrm{cloud}$ and the VMRs of CO and H$_2$O, and the strong degeneracy for cloud deck pressures below one bar is clearly visible. Above this threshold, the VMRs remain constant with decreasing cloud height, because collision-induced absorption (CIA) of H$_2$-H$_2$ and H$_2$-He acts as an additional continuum opacity source, preventing us from probing deeper into the atmosphere. This effect is illustrated in Fig. \ref{Fig_CIA_influence}, showing the relation between line strength and cloud deck pressure for models with and without CIA. At pressures of one bar, the line strength of the model with CIA levels off, and a further increase of $P_\mathrm{cloud}$ does not affect the model.

Previous retrievals of low-resolution observations found no evidence for the existence of clouds on the dayside of WASP-43b. \citet{fraine2021DarkWorldTale} used HST WFC3/UVIS to study the planet's reflected light and were not able to detect the eclipse, leading them to conclude that a significant cloud coverage is unlikely to exist above the probed pressures of around one bar. Phase curve observations with CHEOPS, TESS and HST constrained the maximum cloud deck pressure to a similar value \citep{scandariato2022PhaseCurveGeometric}, and the retrieval of \citet{chubb2022ExoplanetAtmosphereRetrievals} using Spitzer data also found a good fit using cloud-free models. \hbox{\citet{murphy2023LackVariabilityRepeated}} found that, when comparing their models to the Spitzer phase curve, either thin or very thick clouds were favoured. Although they are unable to reject either of the two scenarios due to a limited model resolution and degeneracies, they suggest a relatively cloud-free dayside to be most likely.

Contrary to these results, pure general circulation and cloud formation models of WASP-43b's atmosphere by \citet{venot2020GlobalChemistryThermal}  and \citet{helling2020MineralCloudHydrocarbon} predict several species of clouds at pressures above the detection limit of the aforementioned observational studies, although these results could strongly depend on the initial assumptions and fidelity of the simulations. 

Our retrieval results favour an atmosphere free of clouds down to pressures of at least one bar. In addition, due to the unphysically large CO volume mixing ratios when assuming larger cloud deck pressures, and the absence of observational evidence for clouds in the low-resolution data, we deem it most likely that clouds are either not present or located below this threshold pressure on the dayside. Constraining the posterior distributions to include only MCMC steps with $\log_{10}P_\mathrm{cloud} < 0$, we retrieve a super-solar carbon monoxide abundance of $\log_{10}$(CO) = $-3.0^{+0.8}_{-0.6}$, in agreement with the findings of \citet{irwin20205DRetrievalAtmospheric} and \hbox{\citet{chubb2022ExoplanetAtmosphereRetrievals}}. If the H$_2$O signal is real and not created by spurious noise, the detected feature corresponds to a water abundance of $\log_{10}$(H$_2$O) = $-4.4 \pm$ 0.3. This H$_2$O abundance is consistent within $\sim 1\sigma$ with the results of \mbox{\citet{stevenson2017SpitzerPhaseCurve}} and \citet{welbanks2019DegeneraciesRetrievalsExoplanetary, welbanks2022AtmosphericRetrievalsExoplanets}, but lower than the detections of \citet{kreidberg2014PreciseWaterAbundance} and \citet{irwin20205DRetrievalAtmospheric}. 

We found a broadening of the spectral lines corresponding to an equatorial velocity $v_\mathrm{eq} = 11.7^{+2.8}_{-2.1}\,$km\,s$^{-1}$, which is larger than the rotational velocity $v_\mathrm{rot} = 2\pi R_p / P = 6.29\pm 0.11\,$km\,s$^{-1}$ assuming the planet is tidally locked. 
The difference in velocity can be explained by a high velocity jet causing equatorial super-rotation with a wind speed of $v_\mathrm{jet} = 5.4^{+2.8}_{-2.1}\,$km\,s$^{-1}$. For the rotational profile, we assumed a limb darkening coefficient of $\epsilon = 1$, which is justified by the existence of a hot spot creating a strong contrast between the centre and limb region. Setting this parameter to significantly smaller values does not influence the results by more than 1-2 km\,s$^{-1}$.

The existence of such a jet is supported both by previous observations and simulations. Multiple studies of Spitzer observations  found an eastward shift of the hot spot by $\sim$ 10$^\circ$ \citep[e.g.][]{stevenson2017SpitzerPhaseCurve, chubb2022ExoplanetAtmosphereRetrievals, murphy2023LackVariabilityRepeated}, while \citet{scandariato2022PhaseCurveGeometric} analysed TESS phase curves and found marginal evidence for a $\left(50 ^{+30}_{-20}\right) ^\circ$ eastward phase offset. 
The formation of prograde winds at the equator is generally predicted by 3D general circulation models of tidally locked hot Jupiters (e.g. \citealt{showman2009AtmosphericCirculationHot, rauscher2010ThreedimensionalModelingHot, heng2011AtmosphericCirculationTidally, mayne2014UnifiedModelFullycompressible}). In simulations specifically of WASP-43b, \citet{kataria2015AtmosphericCirculationHot} find that the jet can in principle reach velocities upwards of 4\,km\,s$^{-1}$,  while \citet{zhang2017ConstrainingHotJupiter} report a wind speed of 6.3 km\,s$^{-1}$ at pressure levels associated with CO lines. There has been the notion that both prograde and retrograde streams exist in different regions and heights, and that the retrograde equatorial flow could dominate the overall dynamics \citep{carone2020EquatorialRetrogradeFlow}. However, the follow-up study by \citet{schneider2022ExploringDeepAtmospheres} suggests that the retrograde flow may be an artefact of imposed temperature forcing. In their refined model, the prograde jet emerges as the dominant structure with a velocity of $\sim$\,5\,km\,s$^{-1}$.
Our retrieved equatorial wind speed thus agrees well with previous observations and various simulations of the atmospheric dynamics.

We investigated the influence of smeared spectral lines (see Sect. \ref{Section_Smearing}) on the retrieved $v_\mathrm{eq}$, but we found no significant difference between retrievals with and without smearing. As shown in Fig. \ref{Fig_BroadeningEffects}, the addition of smearing is relevant in the absence of rotational broadening, but only has a marginal effect on the shape of the lines when a strong rotational broadening is present.

\subsection{Determining the C/O ratio}
We conducted a second retrieval, in which the C/O ratio and metallicity [M/H] were set as free parameters instead of the abundances of CO and H$_2$O. We used the \texttt{poor\_mans\_nonequ\_chem} subpackage from petitRADTRANS to calculate chemical equilibrium abundances from C/O ratios and metallicities. Because we concluded previously that there is no evidence for clouds in the probed region of the atmosphere, we did not include a cloud deck in this second retrieval. 

We found C/O = 0.78 $\pm$ 0.09 and [M/H] = 0.2$^{+0.7}_{-0.6}$, while the other parameters ($T$-$p$ profile, $K_p$, $v_\mathrm{max}$, $v_\mathrm{eq}$, $\beta$) were consistent with our initial retrieval with free abundances (see Fig. \ref{Fig_Cornerplot_COFeH}). Calculating the abundances of CO and H$_2$O based on this C/O ratio and metallicity also gave VMRs close to the results of our initial retrieval. According to these calculations, there could be a significant amount of CH$_4$ present ($\log_{10}$(CH$_4) \approx -3.2$) given the retrieved atmospheric conditions, which we were not able to detect with our data. We conducted an injection-recovery test on the data of night 1 with a pure CH$_4$ model of that abundance, and were not able to recover any signal. The line strength of the model had to be increased by a factor of 3 to be able to tentatively detect the injected signal in the data. Therefore CH$_4$ could be present in the atmosphere with an abundance consistent with our retrieved C/O ratio, without being detectable in the data.

Our result is in agreement with the general consensus that the gaseous C/O ratio is significantly smaller than unity \citep{benneke2015StrictUpperLimits, changeat2021ExplorationModelDegeneracies, chubb2022ExoplanetAtmosphereRetrievals}. The retrieved super-solar metallicity matches with findings of \citet[][{[M/H]} = 0.4 - 3.5 $\times$ solar]{kreidberg2014PreciseWaterAbundance} and \citet[][0.4 - 1.7 $\times$ solar]{stevenson2017SpitzerPhaseCurve}, while it is lower than the estimates of \citet[][5 $\times$  solar]{kataria2015AtmosphericCirculationHot}  and \citet[][{[M/H]} = 1.81]{changeat2021ExplorationModelDegeneracies}.

Many previous studies used the Spitzer datasets to constrain the C/O ratio, and these results show a strong dependency on the reduction method and assumed thermal profiles \citep{changeat2021ExplorationModelDegeneracies}. 
The observations presented in this work cover the CO band head in the K band with the high resolution of CRIRES$^+$. Therefore the determination of the CO abundance does not rely on individual data points, as it is informed by a large number of CO absorption lines. Nevertheless, similar to previous works the retrieved C/O ratio is only based on two molecular species, as we were not able to detect other carbon- and oxygen-bearing molecules. This could be either due to small overall atmospheric abundances, or because the observations were not sensitive to the relevant wavelength or pressure regions for these species.

WASP-43b is a target of the JWST Cycle 1 ERS Program \citep[DD-ERS 136, PIs: N. Batalha, J. Bean, K. Stevenson;][]{stevenson2016TransitingExoplanetStudies, bean2018TransitingExoplanetCommunity}, which observed a full phase curve of the planet with JWST/MIRI in the infrared 5-12$\,\mu$m range \citep{venot2020GlobalChemistryThermal}. This observation will be used to study the planetary night side, and to constrain abundances and properties of clouds. A joint retrieval using these not yet published observations with JWST will offer the best bet for an accurate determination of the abundances and the C/O ratio.

\section{Conclusions} \label{Section_Conclusion}
We analysed the thermal emission spectra of WASP-43b observed with the high-resolution spectrograph CRIRES$^+$ by implementing the cross-correlation method. We found a robust CO absorption signal, the first direct detection of this species in WASP-43b's atmosphere, as well as evidence for H$_2$O absorption lines. For this, we removed stellar and telluric contributions using SYSREM, and optimised the number of iterations based on an injection-recovery test. Subsequently, forward-model retrievals were conducted to constrain the atmospheric and orbital parameters of WASP-43b. Our retrieval accounts for the smearing of spectral lines due to the planetary motion during exposures, applies a filter to the model to account for distortion effects introduced by SYSREM, and includes rotational broadening due to the tidally locked rotation and equatorial winds.

We recovered a non-inverted $T$-$p$ profile that agrees well with the equilibrium temperature in the upper regions of the atmosphere. The retrieved CO and H$_2$O abundances are generally consistent with results of previous studies, although there are some studies that report a notably greater H$_2$O VMR. Assuming equilibrium chemistry in the atmosphere, we found a super-solar metallicity and a C/O ratio of 0.78 $\pm$ 0.09. This result is based on the detected signal of CO and H$_2$O only, and does not account for any other carbon- and oxygen-bearing species that were not detectable with our data. In agreement with simulations of tidally locked hot Jupiters, we found a rotational broadening consistent with a fast super-rotating equatorial jet with a speed of $v_\mathrm{jet} = 5.4^{+2.8}_{-2.1}$\,km\,s$^{-1}$. Furthermore, there was no evidence for a significant cloud deck coverage on the dayside above pressures of one bar based on our data.
A combination of these results with the JWST observations of WASP-43b will lead to a larger wavelength coverage and more robust abundance measurements not only of CO and H$_2$O, but potentially also of other carbon- and oxygen-bearing species.

\begin{acknowledgements}
CRIRES$^+$ is an ESO upgrade project carried out by Thüringer Landessternwarte Tautenburg, Georg-August Universität Göttingen, and Uppsala University. The project is funded by the Federal Ministry of Education and Research (Germany) through Grants 05A11MG3, 05A14MG4, 05A17MG2 and the Knut and Alice Wallenberg Foundation. Based on observations collected at the European Organisation for Astronomical Research in the Southern Hemisphere under ESO programmes 108.22PH.004 and 110.249Y.004. F.L. acknowledges the support by the Deutsche Forschungsgemeinschaft (DFG, German Research Foundation) – Project number 314665159. D.C. is supported by the LMU-Munich Fraunhofer-Schwarzschild Fellowship and by the Deutsche Forschungsgemeinschaft (DFG, German Research Foundation) under Germany´s Excellence Strategy – EXC 2094 – 390783311. M.R. and S.C. acknowledge the support by the DFG priority program SPP 1992 “Exploring the Diversity of Extrasolar Planets” (DFG PR 36 24602/41 and CZ 222/5-1, respectively). D.S. acknowledges funding from project PID2021-126365NB-C21(MCI/AEI/FEDER, UE) and financial support from the grant CEX2021-001131-S funded by MCIN/AEI/ 10.13039/501100011033.
\end{acknowledgements}

%
%
\bibliographystyle{aa} 
\bibliography{WASP-43b_References.bib}
\newpage
\appendix

\onecolumn
\section{Observational conditions}

\begin{center}
   \includegraphics[width=\textwidth]{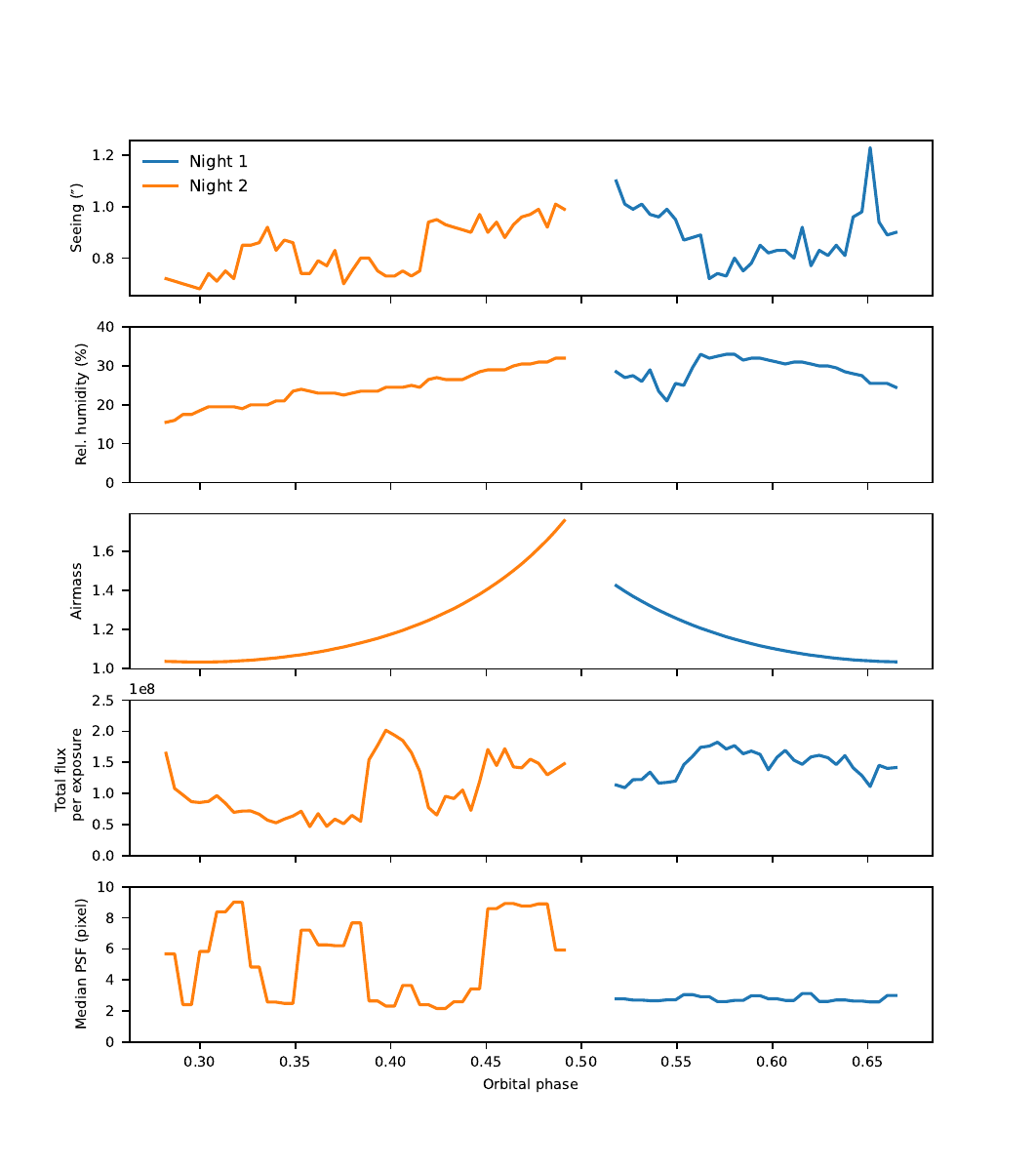}
  \captionof{figure}{Observational conditions for night 1 (blue) and night 2 (orange) as a function of phase. Shown are the seeing (\textit{first panel}), relative humidity (\textit{second panel}), airmass (\textit{third panel}), total flux per exposure (\textit{fourth panel}) and median FWHM of the PSF in spatial direction (\textit{fifth panel}). The worse data quality of night 2 can be contributed to clouds (as seen by the drop in total flux) and the lack of AO causing a significantly larger PSF.}
  \label{Fig_NightlyConditions}
\end{center}
\twocolumn

\clearpage
\section{Additional $K_p$-$v_\mathrm{sys}$ maps}

\begin{figure}[h]
\centering
\includegraphics[width=\hsize]{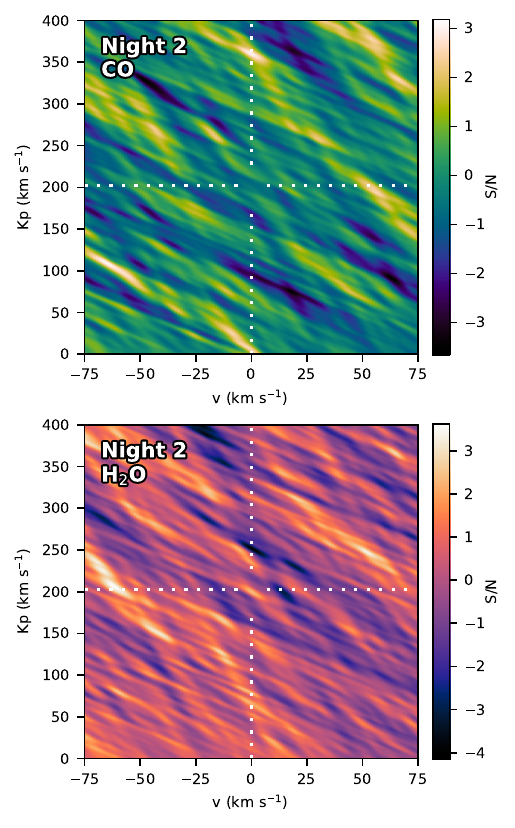}
  \caption{$K_p$-$v_\mathrm{sys}$ map for the CO signal (\textit{top panel}) and H$_2$O signal (\textit{bottom panel}) in night 2 after eight SYSREM iterations. The white dotted lines indicate the expected position of a signal at the literature $K_p$ value in the planetary rest-frame. The absence of a significant signal in either of these species can be attributed to the poor data quality of night 2.}
     \label{Fig_DetMap_Night2}
\end{figure}

\begin{figure}[h]
\centering
\includegraphics[width=\hsize]{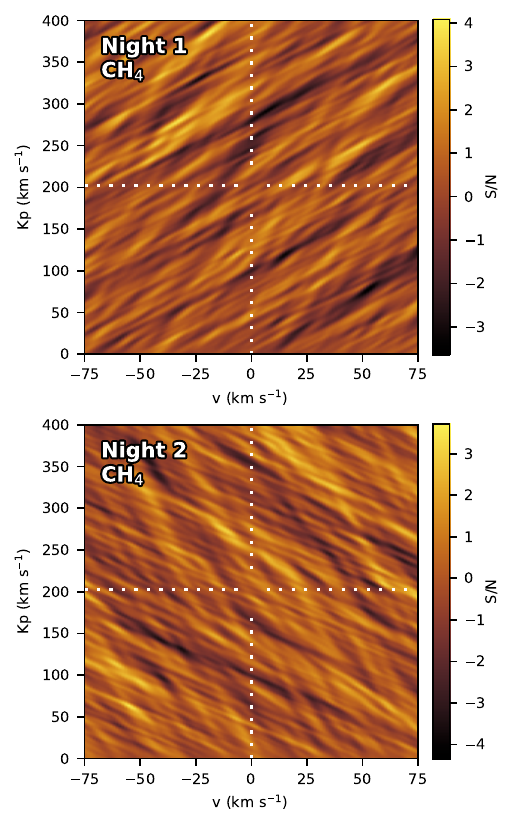}
  \caption{$K_p$-$v_\mathrm{sys}$ map for the CH$_4$ signal in night 1 after ten SYSREM iterations (\textit{top panel}) and night 2 after eight iterations(\textit{bottom panel}). The white dotted lines indicate the expected position of a signal at the literature $K_p$ value in the planetary rest-frame.}
     \label{Fig_DetMap_CH4}
\end{figure}

\clearpage

\onecolumn
\section{Posterior distribution of C/O retrieval}

\begin{center}
  \includegraphics[width=\textwidth]{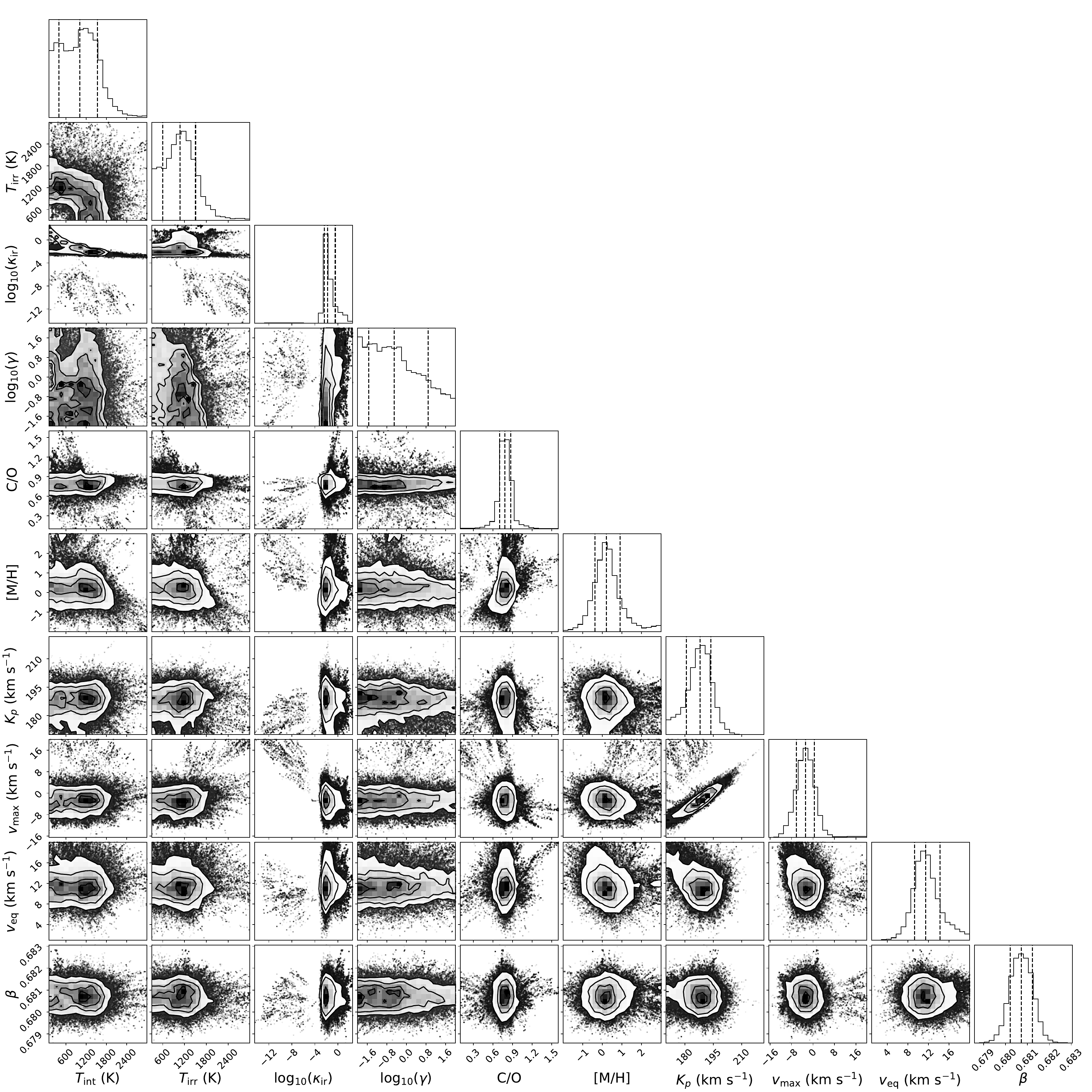}
  \captionof{figure}{Posterior distributions of the retrieval using C/O and [M/H] as free parameters instead of the CO and H$_2$O abundances.}
  \label{Fig_Cornerplot_COFeH}
\end{center}
\twocolumn

\end{document}